\documentclass[12pt,english,super,sort&compress]{elsarticle}
\usepackage{helvet}

\usepackage[T1]{fontenc}
\usepackage[latin9]{inputenc}
\usepackage[letterpaper]{geometry}
\usepackage{textcomp}
\usepackage{amsmath}
\usepackage{amssymb}
\usepackage{graphicx}
\usepackage{setspace}
\makeatletter

\newcommand{\lyxdot}{.}

\usepackage{babel}

\usepackage{epstopdf}

    \def\ps@pprintTitle{%
       \let\@oddhead\@empty
       \let\@evenhead\@empty
       \def\@oddfoot{\reset@font\hfil\thepage\hfil}
       \let\@evenfoot\@oddfoot
    }

\@ifundefined{showcaptionsetup}{}{%
 \PassOptionsToPackage{caption=false}{subfig}}
\usepackage{subfig}
\makeatother

\usepackage{babel}
\begin{document}

\begin{frontmatter}{}

\title{Electrical Double Layer Properties of Spherical Oxide Nanoparticles}

\author{Christian Hunley and Marcelo Marucho\footnote{Email:marcelo.marucho@utsa.edu}}

\address{Department of Physics and Astronomy, The University of Texas at San
Antonio, San Antonio, TX 78249-5003}
\begin{abstract}
The accurate characterization of electrical double layer properties
of nanoparticles is of fundamental importance for optimizing their
physicochemical properties for specific biotechnological and biomedical
applications. In this article, we use classical solvation density
functional theory and a surface complexation model to investigate
the effects of pH and nanoparticle size on the structural and electrostatic
properties of an electrolyte solution surrounding a spherical silica
oxide nanoparticle. The formulation has been particularly useful for
identifying dominant interactions governing the ionic driving force
under a variety of pH levels and nanoparticle sizes. As a result of
the energetic interplay displayed between electrostatic potential,
ion-ion correlation and particle crowding effects on the nanoparticle
surface titration, rich, non-trivial ion density profiles and mean
electrostatic potential behavior have been found. 
\end{abstract}

\end{frontmatter}{}

\begin{keywords} Electrical double layer, Ion distribution, Density
functional theory, Nanoparticles, Surface chemistry \end{keywords}

\section{Introduction}

An increasing number of nanoparticle applications have been recently
proposed in which the pH level of the liquid solution might have a
high impact on their functional behavior\citealp{key-1,key-2-2,key-3-3,key-4-4}.
Consequently, there is a high demand in taking control of the mechanisms
governing this process to optimize the physicochemical properties
of nanoparticles for specific biotechnological and biomedical applications\citep{key-5,key-6,key-7,key-8}.
Having this ability requires an accurate molecular understanding of
the structural and electrostatic properties of liquids surrounding
nanoparticles. A large number of research papers published on colloidal
systems reveal a complex, even not well understood, interplay between
the surface charge density (SCD), Zeta potential (ZP), nanoparticle
size (NS), ionic charge, water density distributions and pH levels
of the surrounding aqueous solution. These articles provide insight
on the formation of electric double layers (EDLs) where the electrostatic
and entropic interactions between the nanoparticle and the aqueous
medium generate a strongly correlated liquid near the surface of the
nanoparticle\citep{key-9,key-10,key-11,key-12,key-13,key-14}. One
major source of complexity in describing EDL properties comes from
the still not fully elucidated role of the pH level on the electrostatic
ion-ion correlation\citep{key-15,key-16}, entropy (e.g. particle
crowding)\citep{key-17} and mean electrostatic potential that contribute
to the ionic driving force\citep{key-18}. Gaining this understating
at the microscopic level is indeed challenging, sometimes impossible,
to obtain using current experimental techniques and conventional mean
field-like Poisson Boltzmann (PB) approaches\citep{key-2,key-3,key-4}
. More sophisticated approaches were proposed to overcome a number
of these limitations \citep{key-19,key-20,key-21,key-22,key-23,key-24,key-25,key-26,key-27,key-28}.
A novel approach named classical solvation density functional theory
(CSDFT) has been recently introduced to study the influence that biological
environments at neutral pH may have on the physicochemical properties
of spherical nanoparticles of different sizes while under a variety
of electrolyte conditions\citep{key-29}. A more recent formulation
of this approach combines CSDFT and surface complexation model (SCM)
\citep{key-30} to account for ion-ion correlation and particle crowding
effects on the surface titration of spherical oxide nanoparticles\citep{key-31}.
The results, successfully validated against experiments, show that
the particle crowding and electrostatic screening effects have a profound
influence on the SCD and ZP of the nanoparticle. 

In this article, we extend the aforementioned work by presenting a
comprehensive analysis to advance the understanding of the role that
pH and nanoparticle size play on the structural and electrostatic
properties of the electrolyte surrounding spherical oxide nanoparticles.
We consider spherical $SiO$ nanoparticles with sizes of different
order of magnitude immersed in a monovalent electrolyte of acid and
alkaline aqueous solutions. We use CSDFT and SCM to calculate the
water and ion density profiles as well as the mean electrostatic potential
(MEP) arising from the charging process generated by the nanoparticle
surface chemistry. Additionally, we calculate the particle crowding,
ion-ion correlation and electrostatic energy contributions to the
ionic potential of mean force (PMF) to identify the dominant interactions
governing the EDL properties. Finally, we compare our results against
Poisson-Boltzmann (PB) predictions to determine the role that the
corrections to the continuum model have in capturing charge inversion,
ionic layering formation, and other important phenomena characterizing
these colloidal systems.

\section{Theory}

In this section we provide the basic formulation of the approach used
in this article. The detailed description can be found in references
\citep{key-29,key-31}.

\subsection{Classical solvation density functional theory for spherical electrical
double layers}

In this approach, we consider a rigid charged spherical nanoparticle
of radius $R$ and uniform surface charge density $\sigma$ surrounded
by an electrolyte solution comprised of $m$ ionic species. We use
the solvent particle model to characterize the electrolyte\citep{key-31}.
Each ionic species $i$ is represented by bulk Molar concentration
$[\rho_{i}^{0}]$, a charged hard sphere of diameter $d_{i}$, and
total charge $q_{i}=ez_{i}$, where $e$ is the electron charge and
$z_{i}$ is the corresponding ionic valence (see Fig. 1(b)). Additionally,
the solvent molecules are represented as a neutral ion species whereas
the solvent electrostatics is considered implicitly by using the continuum
dielectric environment with a dielectric constant $\epsilon=78.5$.
The nanoparticle-liquid interaction induces inhomogeneous ion profiles
$\left[\rho_{i}(r)\right]$ which are calculated using CSDFT as follows:

\begin{align}
\left[\rho_{i}(r)\right]=\Bigg\{ & \begin{aligned}\left[\rho_{i}^{0}\right]exp\{\Delta E_{i}(r,\{[\rho_{j}])\}, & r>R+d_{i}/2\\
0, & r\leq R+d_{i}/2
\end{aligned}
\label{eq:ionprofilefinal}
\end{align}
where $\Delta E_{i}(r,\{[\rho_{j}])\equiv-\beta q_{i}\psi(r,\{[\rho_{j}]\})+\Delta c_{i}^{(1)hs}(r;\{[\rho_{j}]\})+\Delta c_{i}^{(1)res}(r;\{[\rho_{j}]\})$
stands for the ionic PMF per unit of thermal energy $KT$, $\beta=1/kT$
, $k$ is the Boltzmann constant, $T$ the temperature (=298.15K),
and $c_{i}^{(1)hs}(r;\{[\rho_{j}]\})$ and $c_{i}^{(1)res}(r;\{[\rho_{j}]\})$
are the hard sphere (particle crowding) and residual electrostatic
ion-ion correlation functions, respectively. $\psi(r,\{[\rho_{j}]\})$
represents the MEP for spherical nanoparticles

\begin{equation}
\psi(r,\{[\rho_{j}]\})=\frac{e}{\mbox{\ensuremath{\epsilon}}}\int_{r}^{\infty}\frac{dr'}{r'^{2}}\left\{ \frac{4\pi R\sigma}{e}+4\pi\int_{0}^{r'}dr'r'^{2}\sum_{i}z_{i}\rho_{i}(r')\right\} \label{eq:psi_def}
\end{equation}
which is the formal solution of the PB equation for an homogeneous
anisotropic dielectric media $\epsilon$

\begin{equation}
\begin{array}{c}
\nabla^{2}\psi(r,\{[\rho_{j}]\})=-\frac{1}{\epsilon}\sum_{i=1}^{m}z_{i}\left[\rho_{i}(r)\right]\}\\
\\
\epsilon\partial\psi(r,\{[\rho_{j}]\})/\partial r|_{r=s}=-\sigma,\qquad\psi(r,\{[\rho_{j}]\})|_{r\rightarrow\infty}\rightarrow0,
\end{array}\label{eq:poissoneq}
\end{equation}
 with the surface charge layer position defined as $s\equiv R+<\{d_{i}\}>$
and $<\{d_{i}\}>\equiv N_{A}l_{B}^{3}\sum_{i}z_{i}^{2}[\rho_{i}^{0}]d_{i}/(2m)$.
In the later definition $N_{A}$ and $l_{B}$ stand for the Avogadro
number and the Bjerrum length, respectively.

\begin{figure}
\begin{centering}
\subfloat[Poisson\_Boltzmann (PB) model]{\includegraphics[angle=90,scale=0.42]{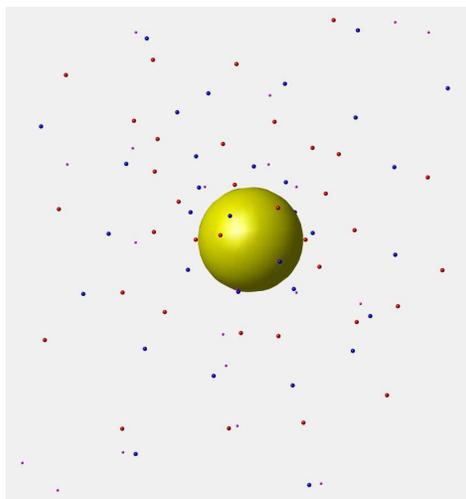} }$\quad$$\quad$\subfloat[Classical Solvation Density Functional (CSDFT) Model ]{\includegraphics[scale=0.45]{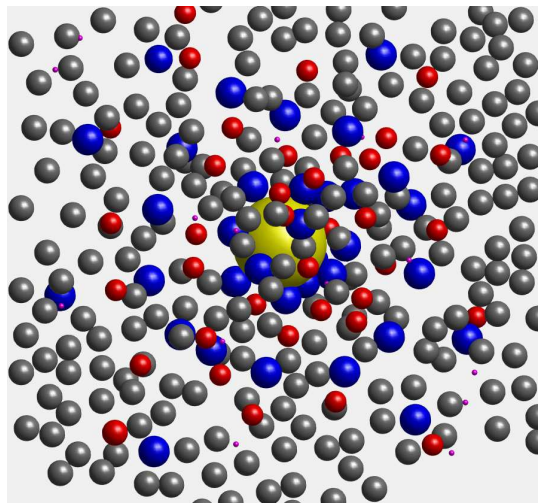} }
\par\end{centering}

\caption{\label{fig:PB-CSDFT}In both models the Silica nanoparticle is displayed
at the center of the figure in yellow color whereas \textit{Na}$^{+}$,\textit{Cl}$^{-}$
and \textit{H}$^{+}$ ions surrounding the nanoparticle are displayed
in blue, red and magenta colors, respectively. Additionally, water
molecules are displayed in gray color in the CSDFT model. In both
models the Silica nanoparticle is represented as a hard sphere with
uniform charge density on the surface whereas ions are represented
by point-like charged particles in PB model and charged hard spheres
with crystalline size in CSDFT model\citep{key-24}. In both models
the electrostatics of water is treated implicitly by using the continuum
dielectric, but CSDFT adittionally includes the water steric interaction
explicitly by considering the water molecules as a neutral hard sphere
(uncharged ions) at experimental size and concentration. }
\end{figure}

\subsection{Surface complexation model}

In order to account for the titration that regulates the nanoparticle
surface charge density $\sigma$ we consider the following two protonation
reactions of single SO-coordinated sites :

\begin{equation}
SOH\leftrightarrow SO^{-}+H^{+},\qquad SOH+H^{+}\leftrightarrow SOH_{2}^{+}.\label{eq:chemicalreactions}
\end{equation}
with equilibrium constants $K_{A}$ and $K_{B}$ 

\begin{equation}
K_{A}=\frac{N_{SO^{-}}[H^{+}]_{s}}{N_{SOH}},\qquad K_{B}=\frac{N_{SOH_{2}^{+}}}{N_{SOH}[H^{+}]_{s}}.\label{eq:equilibriumconstants}
\end{equation}

In the above expressions, $N_{SOH}$, $N_{SO^{-}}$, and $N_{SOH_{2}^{+}}$
are the surface site densities of $SOH$, $SO^{-}$ and $SOH_{2}^{+}$,
respectively. $[H^{+}]_{s}$ is the concentration of $H^{+}$ ions
at the SCD position $s$, namely 

\begin{equation}
[H^{+}]_{s}\equiv\left[\rho_{H}(s)\right]=[\rho_{H}^{0}]exp\{\Delta E_{H}(s,\{[\rho_{j}])\},\label{eq:protonation}
\end{equation}
where $\Delta E_{H}(s,\{[\rho_{j}])=-\beta\zeta+\Delta c_{H}^{(1)hs}(s;\{[\rho_{j}]\})+\Delta c_{H}^{(1)res}(s;\{[\rho_{j}]\})$
is the hydrogen PMF per unit of thermal energy $KT$, $\zeta\equiv\psi(s,\{[\rho_{j}]\})$
is the ZP, and $\Delta c_{H}^{(1)hs}$ and $\Delta c_{H}^{(1)res}$
represent the hydrogen hard sphere (particle crowding) and ion-ion
correlation contributions, respectively. The bulk concentration of
$H^{+}$ ions is represented by $[\rho_{H}^{0}]$, which is related
to the $pH$ value of the bulk liquid at infinity dilution by the
expression $pH=-Log([\rho_{H}^{0}])$. The total number site density
of functional groups on the SCD position is $N_{total}=N_{SO^{-}}+N_{SOH}+N_{SOH_{2}^{+}}$
and the SCD is $\sigma=-F(N_{SO^{-}}-N_{SOH_{2}^{+}})$, where $F$
represents the Faraday constant . Writing the densities sites in terms
of the equilibrium constants (\ref{eq:protonation}), the SCD can
be calculated as follows:

\begin{equation}
\sigma=-FN_{total}\frac{K_{A}-K_{B}\left[\left[\rho_{H}(r)\right]_{r=s}\right]^{2}}{K_{A}+\left[\rho_{H}(r)\right]_{r=s}+K_{B}\left[\left[\rho_{H}(r)\right]_{r=s}\right]^{2}}\label{eq:complexation}
\end{equation}

Note that expression (\ref{eq:complexation}) describes the effects
of the structural and electrostatic properties of the electrolyte
on the SCD, whereas the boundary condition in expression (\ref{eq:poissoneq})
accounts for the SCD effects on the structural and electrostatic properties
of the electrolyte. Therefore, eqns (\ref{eq:ionprofilefinal})-(\ref{eq:complexation})
must be solved self-consistently.

\subsection{Electrolyte solutions and surface titration characterization}

In this work, we utilize the same parameters used in our previous
work\citep{key-31}. We consider silica oxide nanoparticles with total
density number of active functional group $N_{total}=2*10^{-6}mol/m^{2}$
and equilibrium constants $pK_{A}=\text{\textminus}log(K_{A})=6.8$
(protonation) and $pK_{B}=\text{\textminus}log(K_{B})=1.7$ (deprotonation).
We consider a single salt comprised of mono-valent ions ($NaCl$).
The $pH$ of the solution is adjusted by adding $NaOH$ and $HCl$
solutions to the electrolyte. The resulting electrolyte solutions
contain five ion species ($Na^{+}$, $Cl{}^{-}$, $H{}^{+}$, $OH^{-}$,
$H_{2}O$). The free proton and hydroxyl ion bulk concentrations are
given by the well-known expressions $[\rho_{H}^{0}]=10^{-pH},\quad and\quad[\rho_{OH}^{0}]=10^{-(14-pH)},$
respectively, and the bulk concentrations of the electrolyte are chosen
to satisfy the bulk electroneutrality condition.

\subsection{PB approach}

PB is a particular case of the CSDFT approach. Indeed, the expressions
introduced in sections 2.1-2.3 for CSDFT recover the (non linear)
PB approach by setting all ion sizes equal to zero (see Fig. 1(a)).
In particular, $s=R$, $\Delta c_{i}^{(1)res}(r;\{[\rho_{j}]\})=0,$
and $\Delta c_{i}^{(1)hs}(r;\{[\rho_{j}]\})=0$ in continuum models.

\section{Results and discussion}

In this section, we present and discuss the results predicted by CSDFT
to elucidate the role that pH and nanoparticle size play on the EDL
properties of silica oxide nanoparticles. Subsequently, we compare
CSDFT and PB results to remark the physics going beyond continuum
models. Note the Figures showing dashed and solid lines represent
CSDFT and PB predictions, respectively.

\subsection{CSDFT Results}

\begin{figure}
\subfloat[]{\includegraphics[scale=0.33]{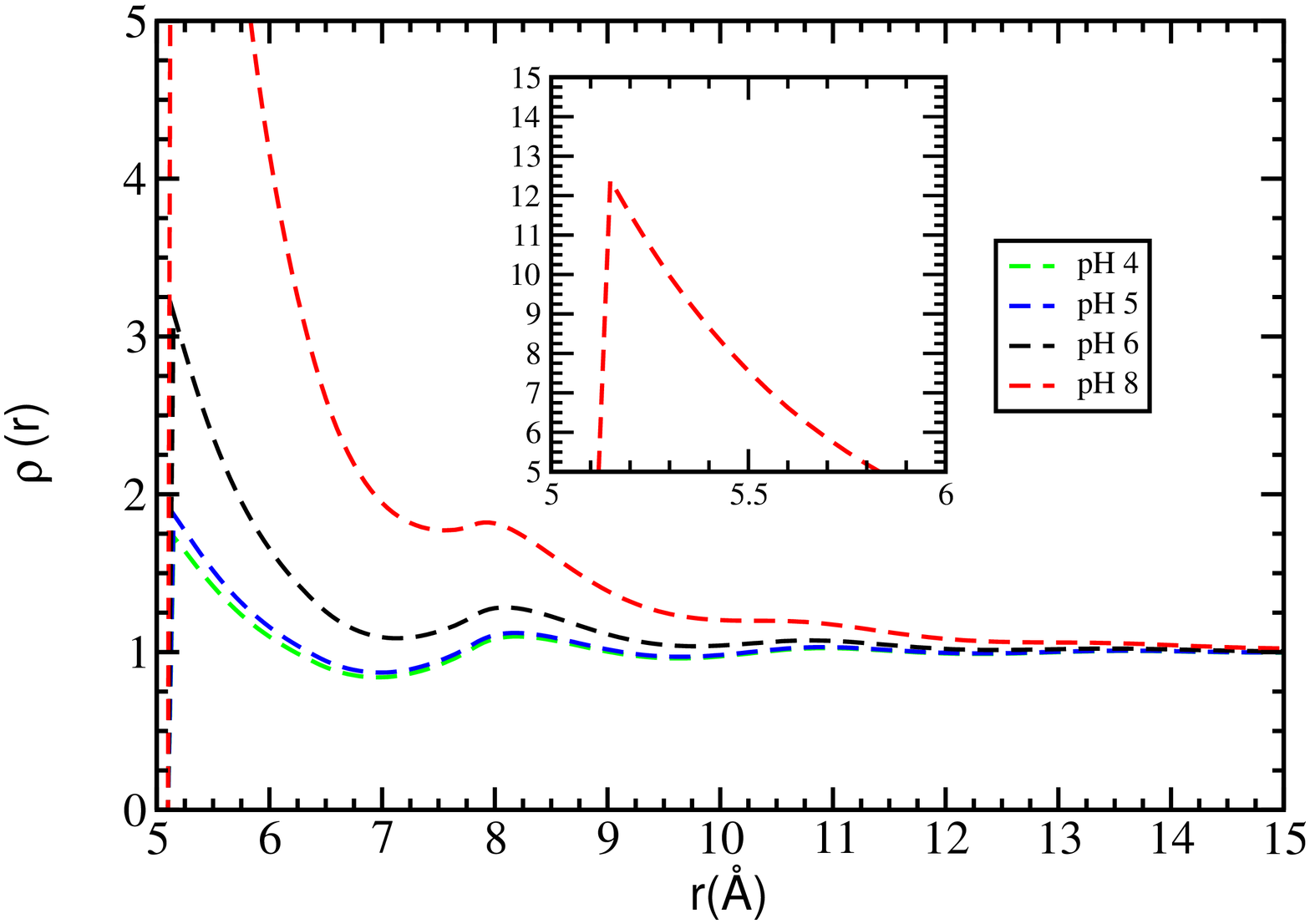}}\subfloat[]{\includegraphics[scale=0.33]{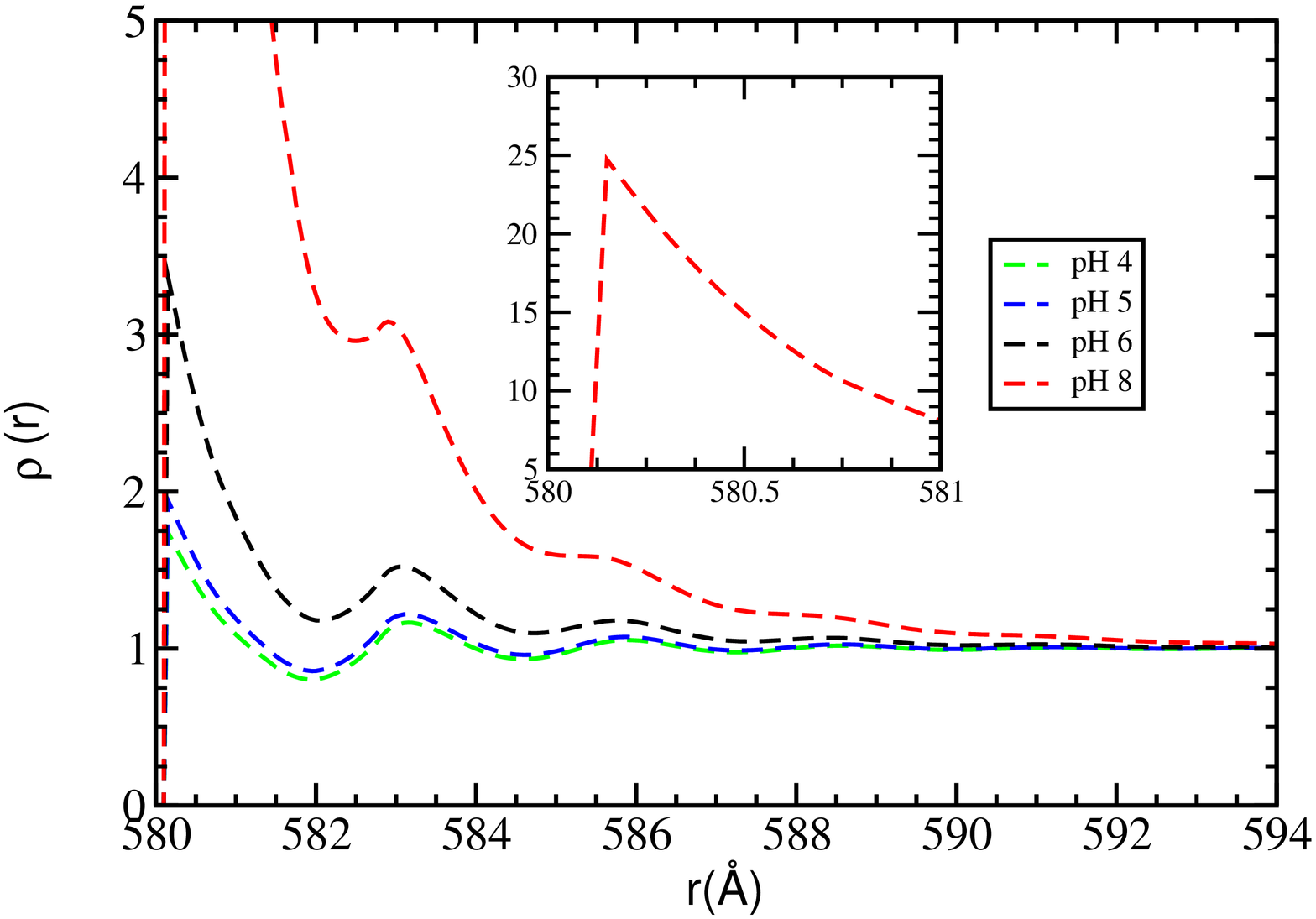}}\caption{\label{fig:Hydrogen-profiles-SPM-1}Hydrogen density profiles per
bulk concentration predicted by CSDFT for 0.8$M$ $NaCl$ salt concentration.
Green, blue, black and red represents the Hydrogen density profiles
for pHs 4, 5, 6,and 8, respectively. Plot (a) corresponds to 5Å nanoparticle
size whereas plot (b) corresponds to 580Å nanoparticle size.}
\end{figure}

\begin{figure}
\subfloat[]{\includegraphics[scale=0.33]{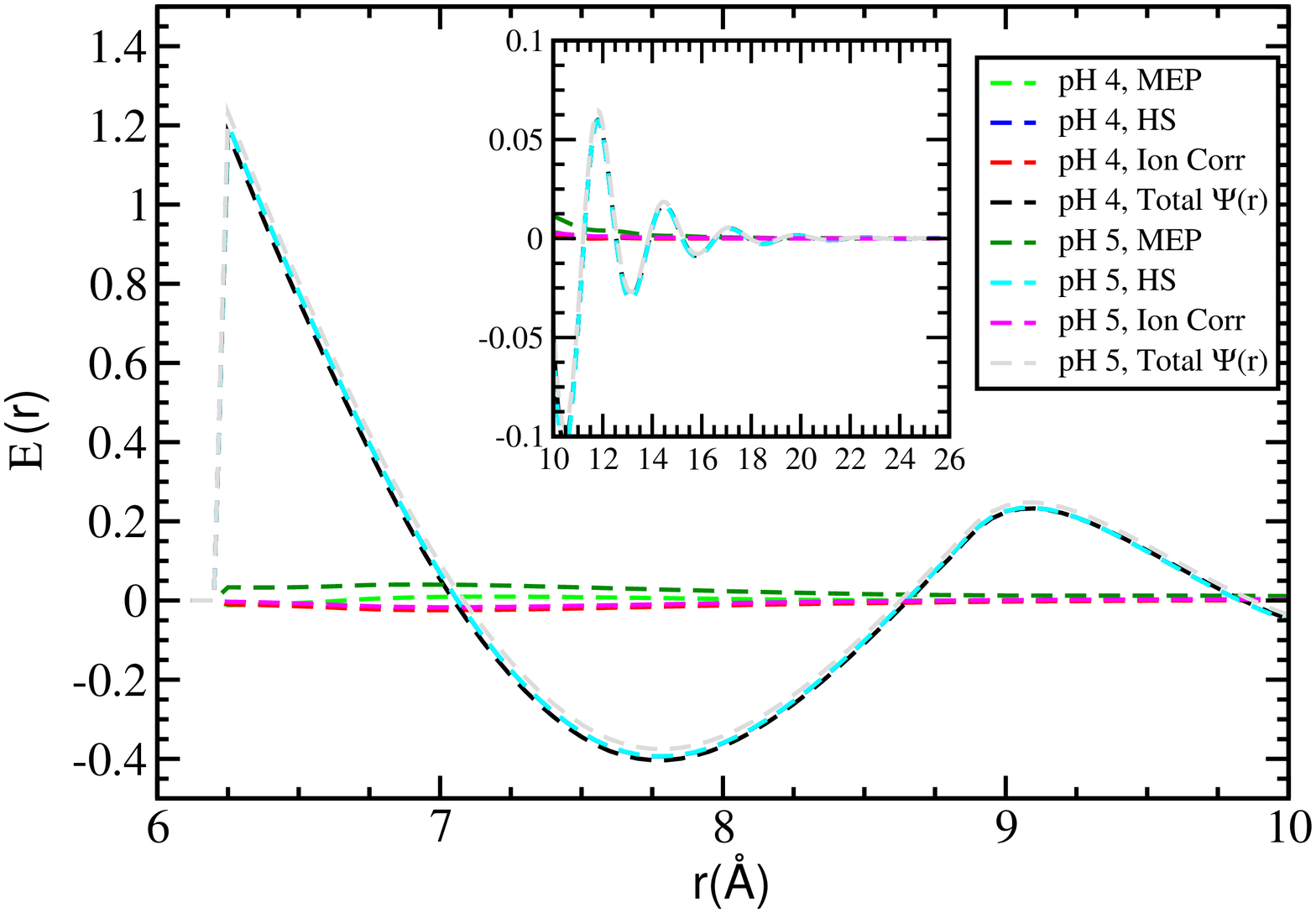}}\subfloat[]{\includegraphics[scale=0.33]{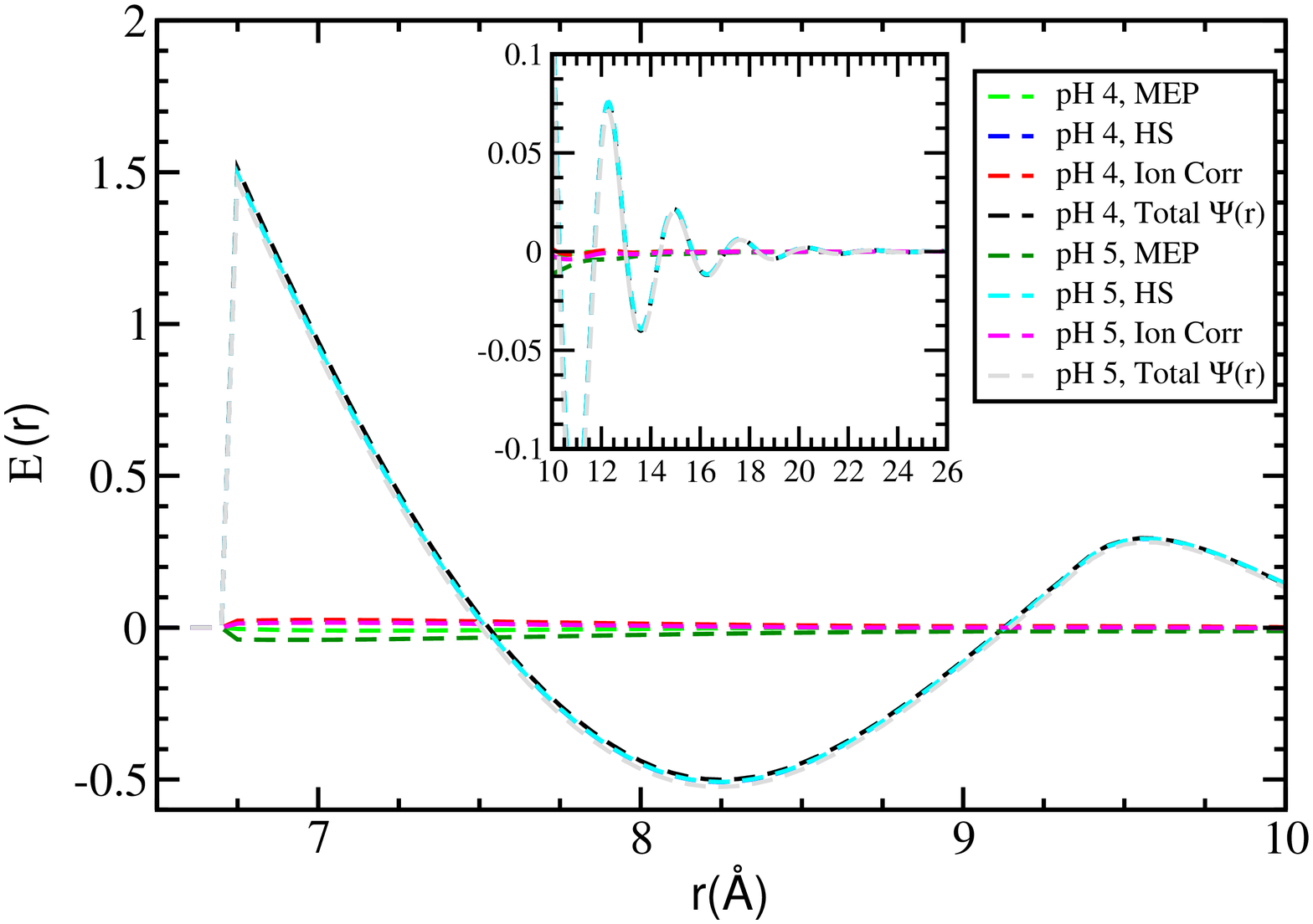}}

\subfloat[]{\includegraphics[scale=0.33]{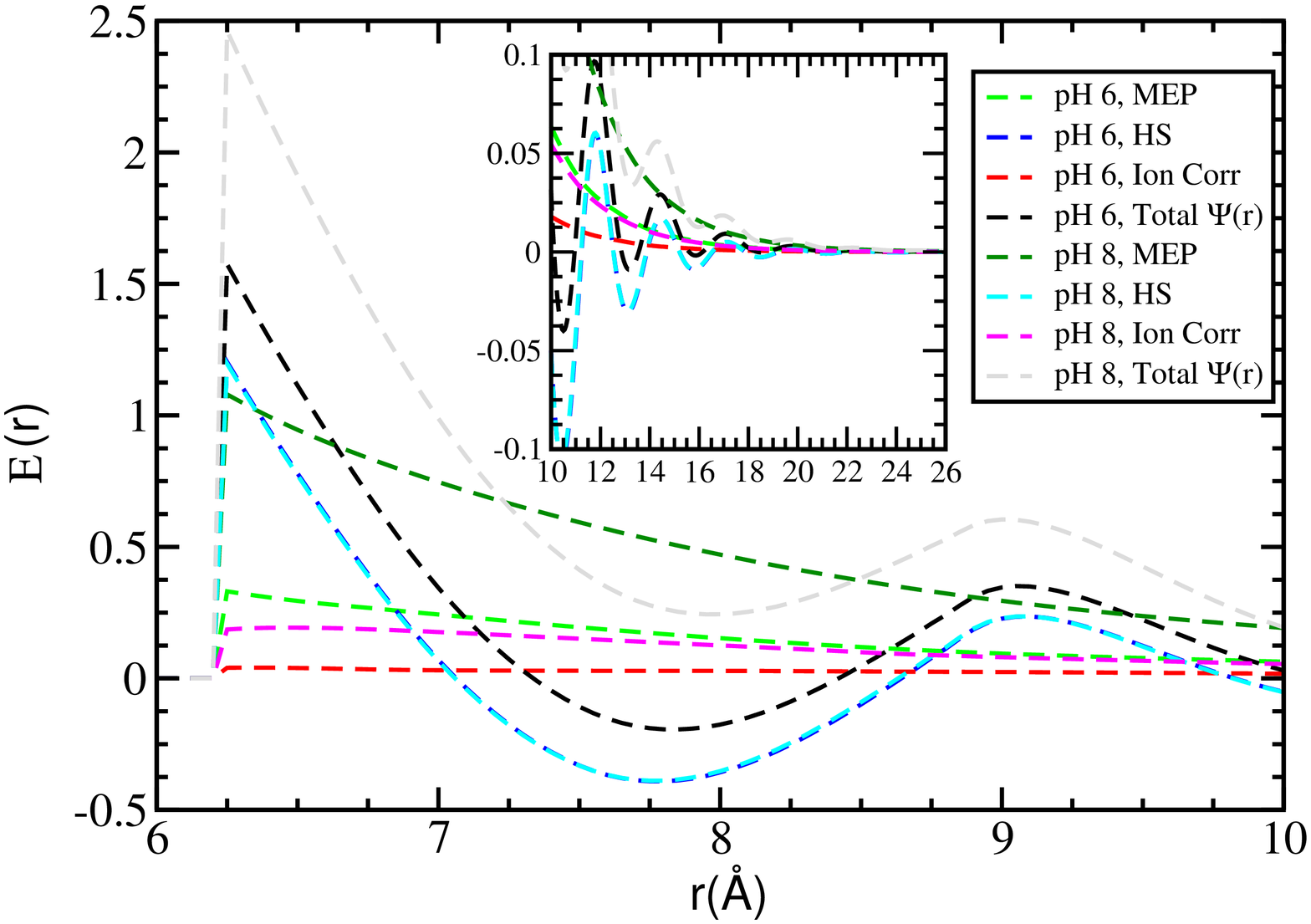}}\subfloat[]{\includegraphics[scale=0.33]{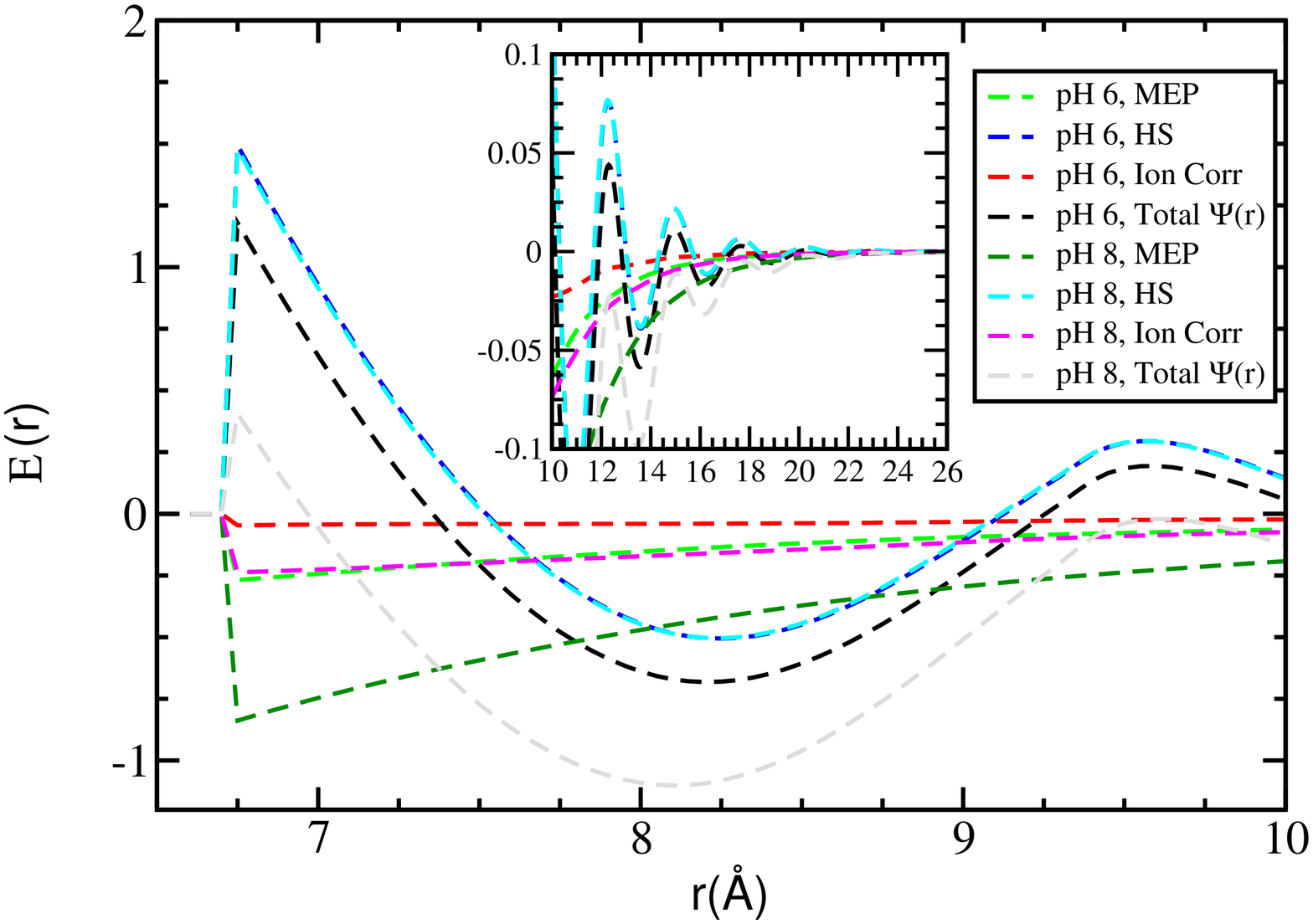}}\caption{\label{fig:ion-contributions-5A-1} $E(r)$ represents the mean electrostatic
potential (=MEP), entropy (=HS), ion-ion correlation (=Ion Correl)
and ionic potential of mean force (=Total $\Psi$) energies per unit
of thermal energy $KT$ predicted by CSDFT for a 0.8$M$ $NaCl$ salt
concentration and 5Å nanoparticle size. Light green and dark green
colors correspond to MEP, blue and cyan colors correspond to entropy,
red and magenta colors correspond to ion-ion correlation, and black
and gray colors correspond to the ionic potential of mean force. Plots
(a) and (b) represent the effects of low pHs (4 and 5) on $Na^{+}$
and $Cl^{-}$ ions, respectively, whereas plots (c) and (d) are those
corresponding to the effects of high pHs (6 and 8).}
\end{figure}

\begin{figure}
\subfloat[]{\includegraphics[scale=0.33]{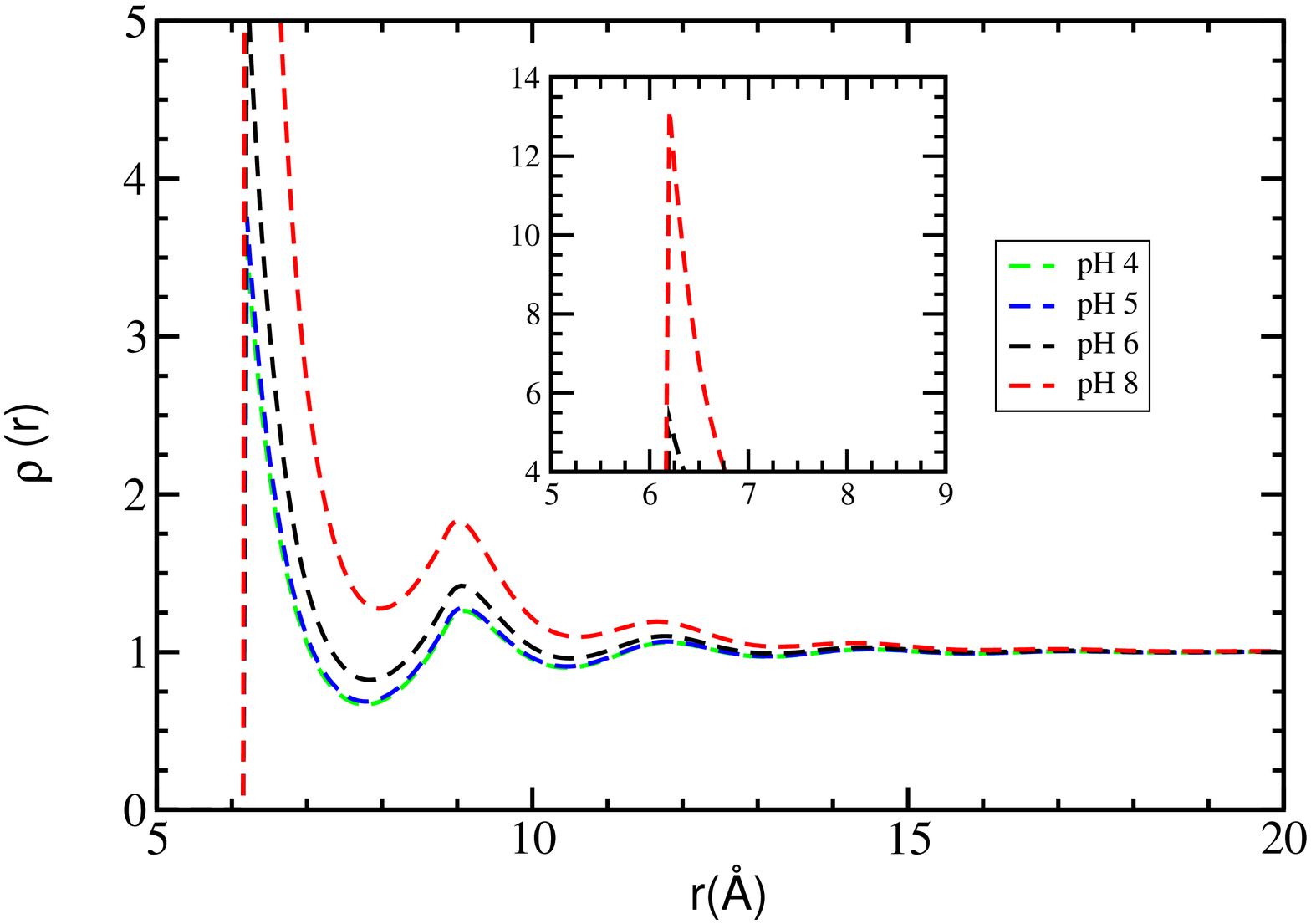}}\subfloat[]{\includegraphics[scale=0.33]{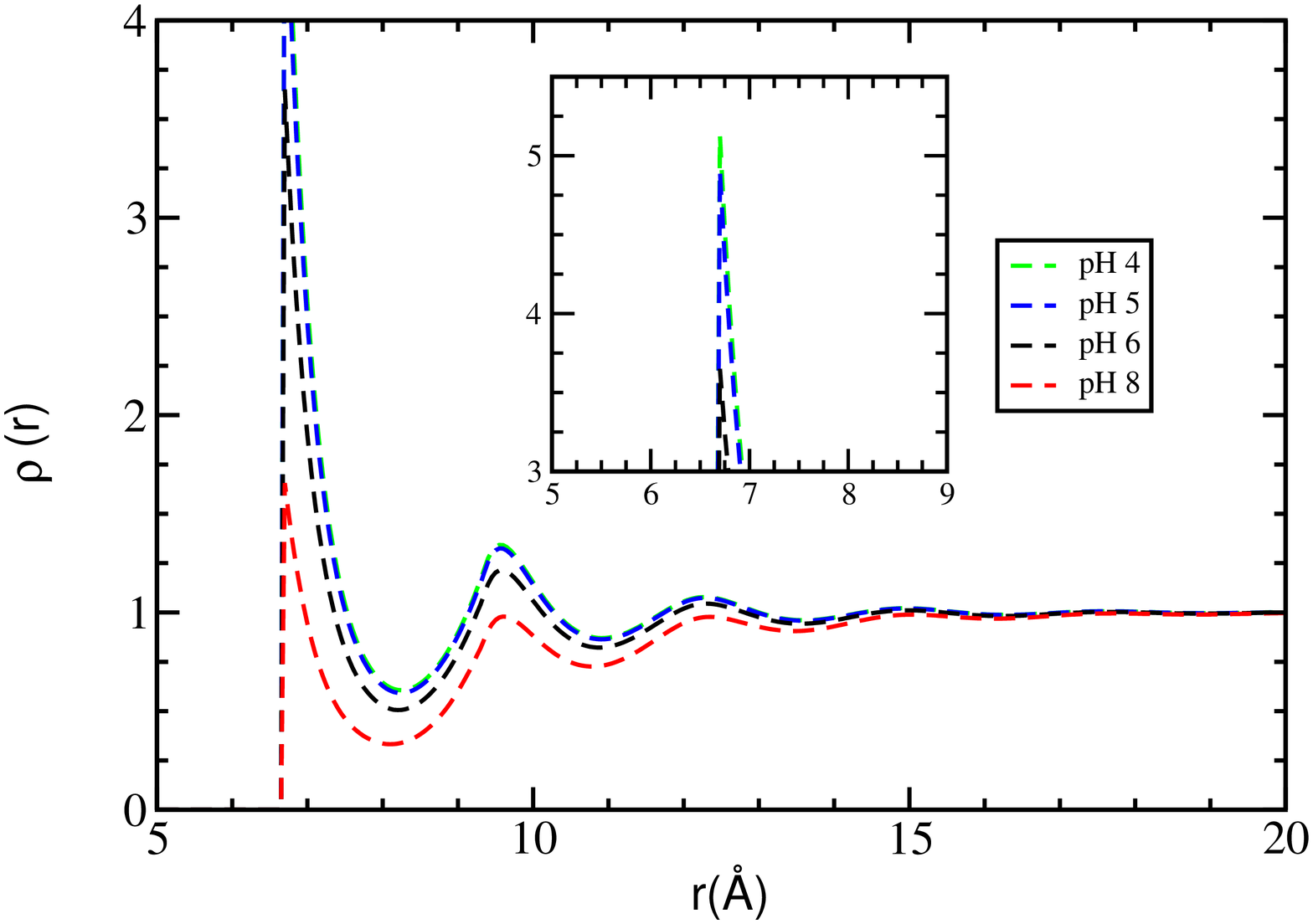}}

\subfloat[]{\includegraphics[scale=0.33]{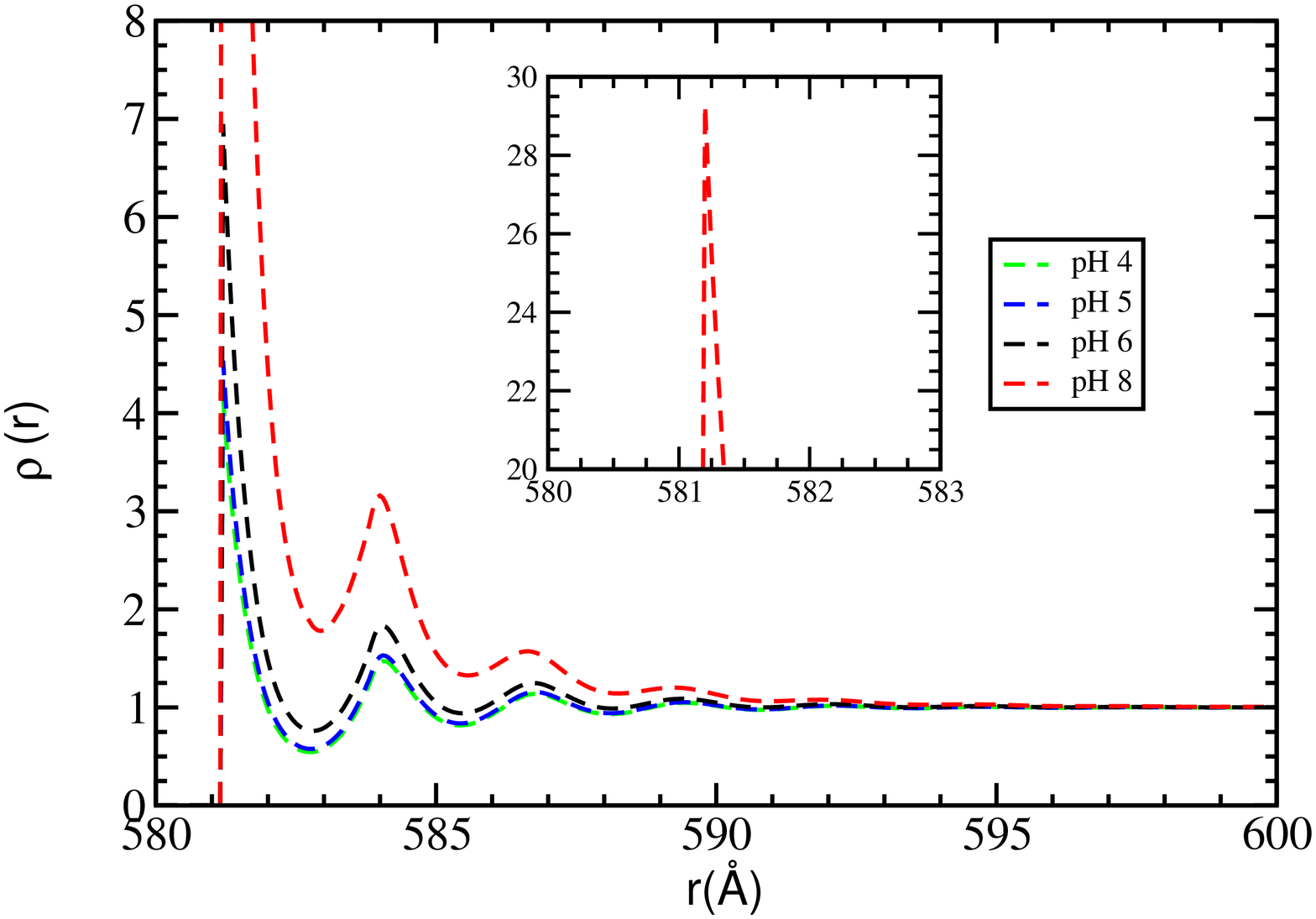}}\subfloat[]{\includegraphics[scale=0.33]{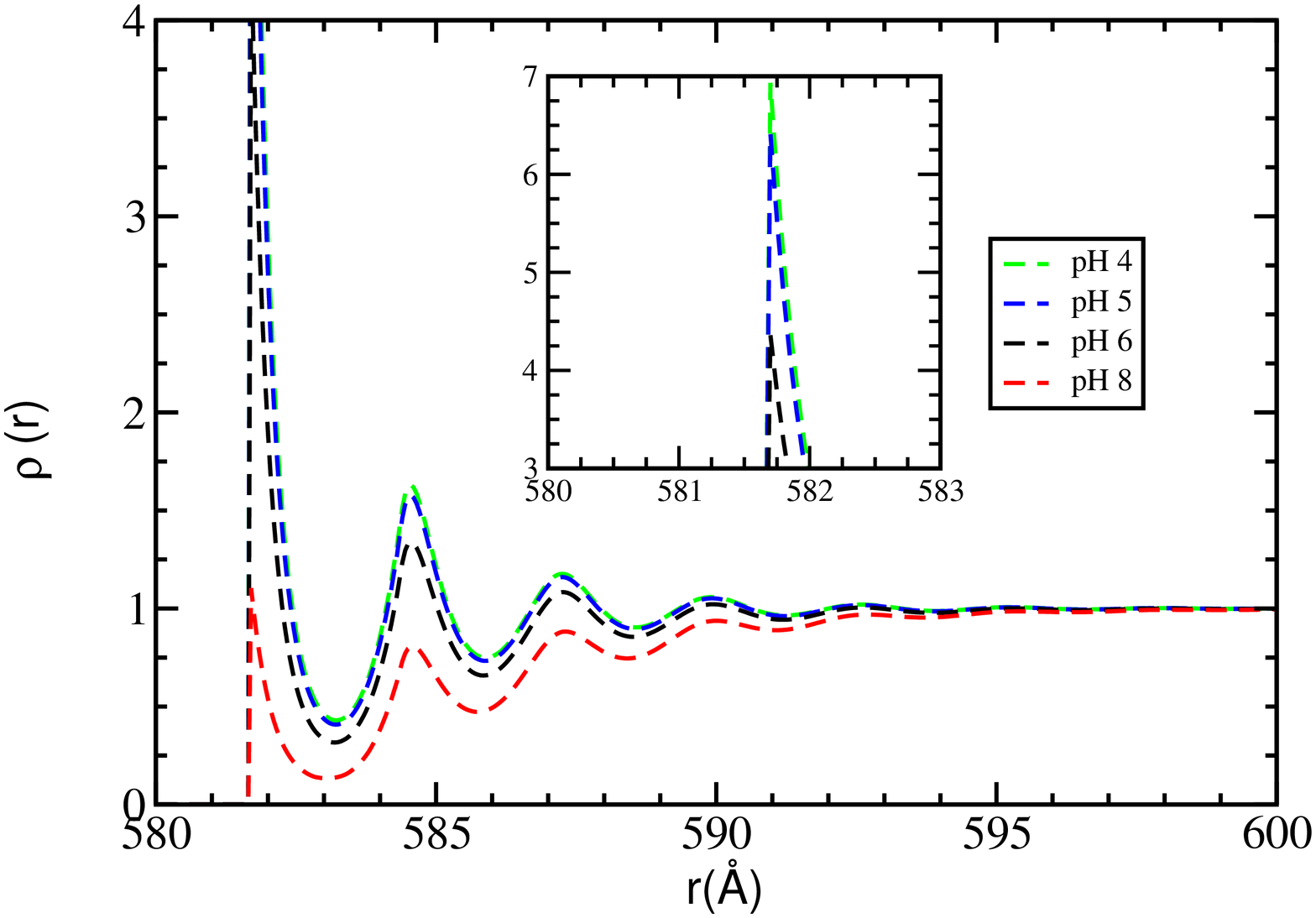}}\caption{\label{fig:pH-effects-on-profiles-5A-1}Ion density profiles per bulk
concentration predicted by CSDFT for 0.8$M$ $NaCl$ salt concentration,
5Å and 580Å nanoparticle sizes, and several pH levels. Green, blue,
black and red colors represent ion density profiles for pHs 4, 5,
6, and 8, respectively. Plot (a) and (c) show $Na^{+}$ density profiles
whereas plots (b) and (d) show $Cl^{-}$ density profiles. Plots (a)-(b)
and (c)-(d) correspond to 5Å and 580Å nanoparticle sizes, respectively. }
\end{figure}

\begin{figure}
\subfloat[]{\includegraphics[scale=0.33]{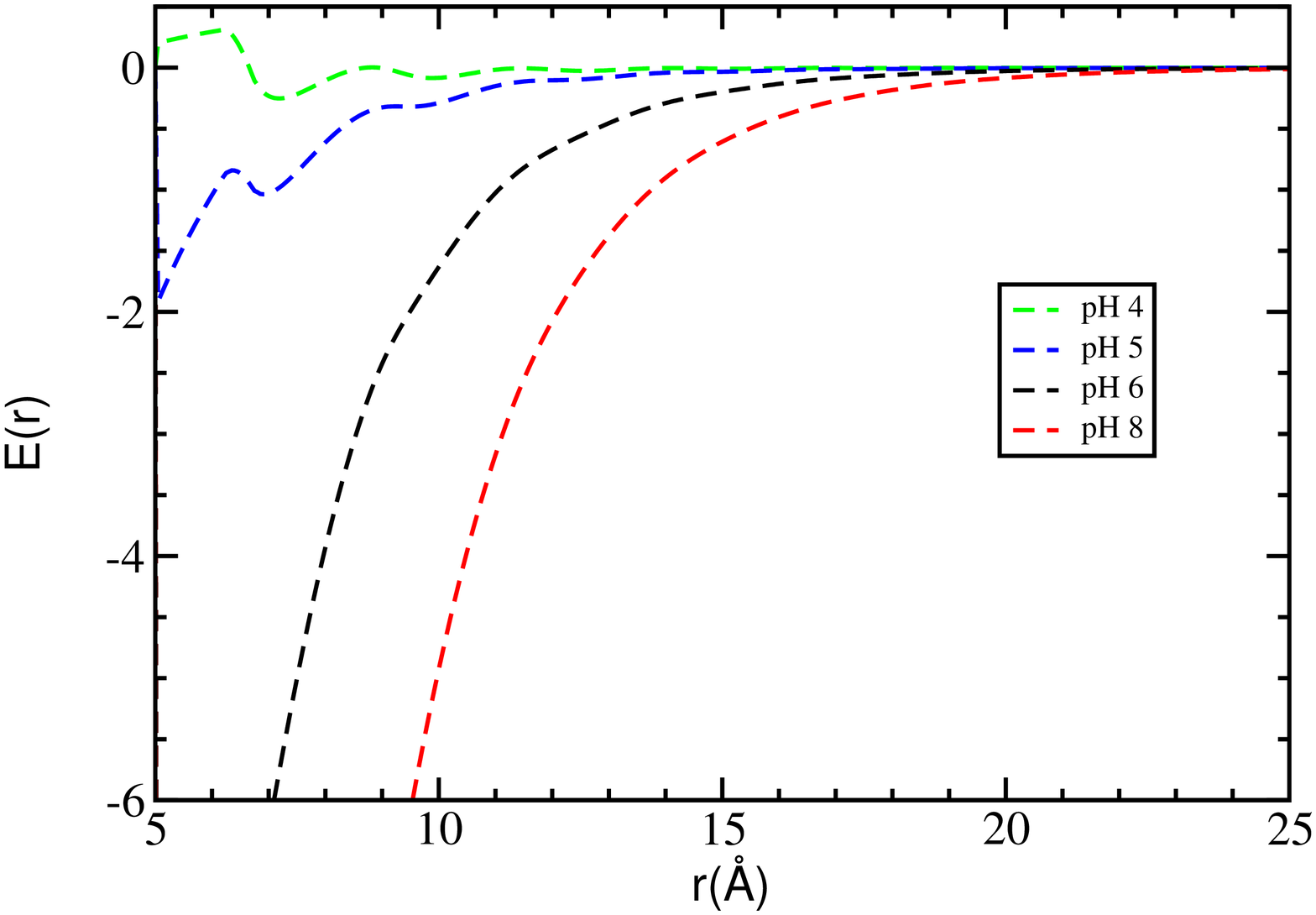}}\subfloat[]{\includegraphics[scale=0.33]{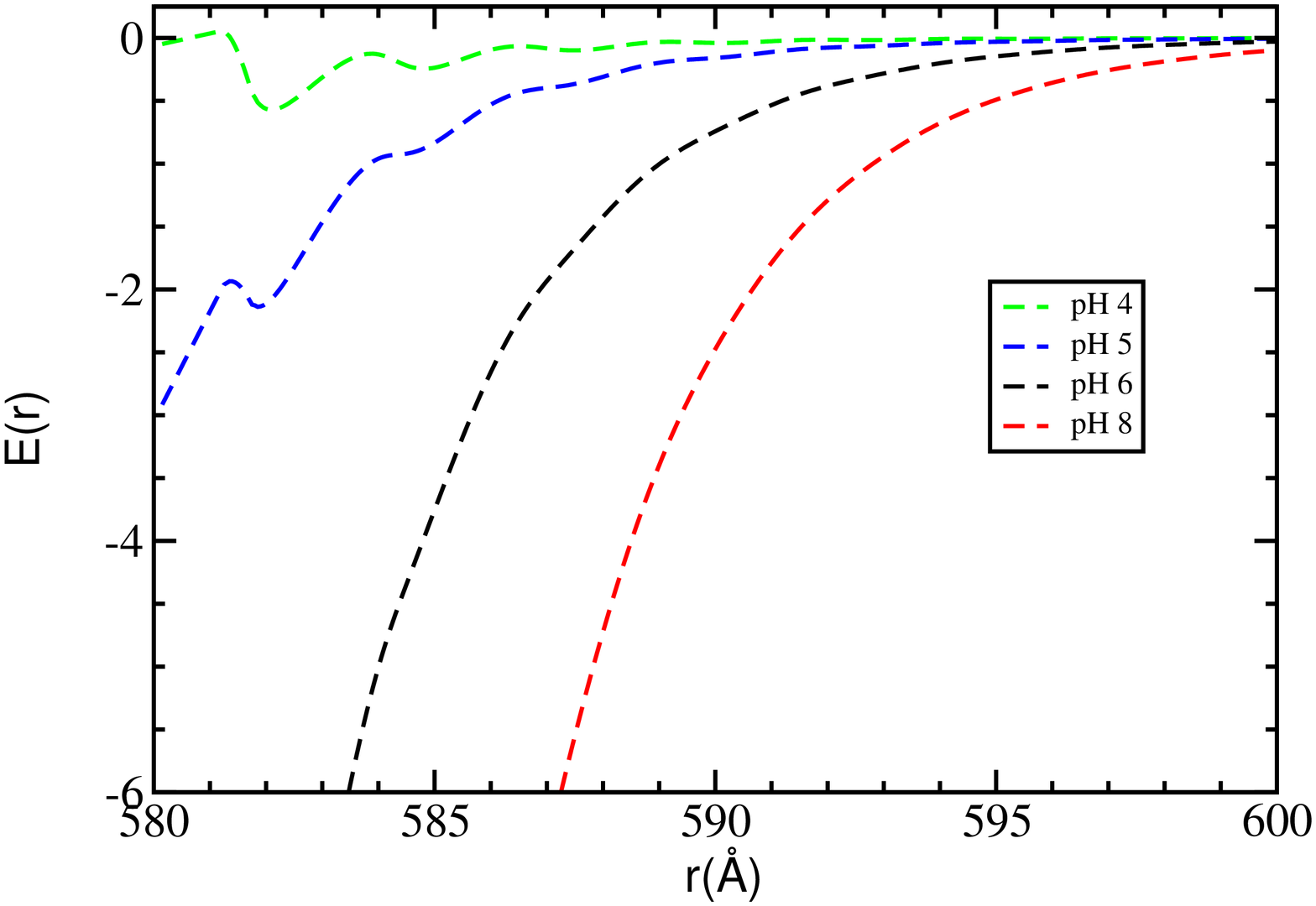}}\caption{\label{fig:MEP-SPM-1}Mean electrostatic potential (MEP) per unit
of thermal energy $KT$ predicted by CSDFT for 0.8$M$ $NaCl$ salt
concentration. Green, blue, black and red represents the MEP for pHs
4, 5, 6,and 8, respectively. Plots (a) corresponds to 5Å nanoparticle
size whereas plot (b) corresponds to 580Å nanoparticle size.}
\end{figure}

In our first analysis we study a 5Å nanoparticle size in 0.8$M$ monovalent
electrolyte solution. Overall, our results show a high impact of the
pH level on the structural and electrostatic properties of spherical
electrical double layers. This behavior comes from expression (\ref{eq:equilibriumconstants})
which explicitly establishes the influence of pH level and equilibrium
constants on the number of active functional groups deprotonated ($N_{SO^{-}}$)
and protonated ($N_{SOH_{2}^{+}}$) on the surface, namely

\begin{equation}
\begin{array}{c}
N_{SO^{-}}=N_{SOH}10^{-6.8+pH}exp(-\Delta E_{H})\\
N_{SOH_{2}^{+}}=N_{SOH}10^{-1.7-pH}exp(\Delta E_{H})
\end{array}\label{eq:charging_mechanisim}
\end{equation}
where $exp(\Delta E_{H})$ represents the normalized hydrogen density
evaluated at the surface charge position $s$. This value can be estimated
by the height of the first peak in the curves in Fig. \ref{fig:Hydrogen-profiles-SPM-1}
(a). Therefore, at low pH levels (e.g. 4 to 5), expression (\ref{eq:charging_mechanisim})
predicts that the density number of active functional groups deprotonated
are partially compensated by those protonated on the nanoparticle
surface. This generates a poor charging mechanism which induces small
surface charge densities (see eqn (\ref{eq:complexation})) and generate
weak nanoparticle-liquid electrostatic interactions. This phenomenon
is shown in Fig. \ref{fig:ion-contributions-5A-1} on the counter-ionic
and co-ionic PMF (dashed black and gray lines), where there is a lowering
of the electrostatic potential energy (dashed light green and dark
green lines) and the ion-ion correlation (dashed red and pink lines)
contributions whereas the ionic entropy (dashed blue and cyan lines)
contribution remains unchanged. According to this analysis, we arrived
at the conclusion that the driving force governing the structural
and electrostatic properties of the EDL in acid electrolyte solutions
comes from the ionic entropy energy. This behavior of the ionic PMF
generates ion density profiles characterized by an increase of the
accumulation of co-ions ($Cl^{-}$) and depletion of counter-ions
($Na^{+}$) near the surface of the nanoparticle compared to those
in bulk concentrations (Fig. \ref{fig:pH-effects-on-profiles-5A-1}
(a)-(b)). Interestingly, there is a higher accumulation of co-ions
than counter-ions found in the first shell at low pH levels due to
the ionic asymmetry size effects. Indeed, even though there is the
same co-ions and counter-ions bulk concentration (=0.8$M$), the size
of the co-ions are larger than the counter-ions which induces larger
entropic contributions to the ionic PMF (see dashed blue and cyan
lines in Fig. \ref{fig:ion-contributions-5A-1}). For instance, we
find around 44 \% more $Cl^{-}$ than $Na^{+}$ ions in the first
shell at pH level 4. On the other hand, our results on MEP at low
pH levels reveal charge inversion around a separation distance of
1.2 Å from the nanoparticle surface (Fig. \ref{fig:MEP-SPM-1} (a)).
This is caused by an excess in counter-ion contributions to the MEP
which generates a charge screening that overcompensates the electrostatic
potential produced by the small nanoparticle charge. Additionally,
the MEP near the nanoparticle surface is characterized by a non-monotonic
short-range behavior. This non-trivial and counter-intuitive behavior
of the MEP comes from expression (\ref{eq:psi_def}) which explicitly
establishes the role that pH plays on the surface charge density (first
term) and ion density profiles (second term). At low pH levels, the
first term becomes small whereas the second term provides the nontrivial
behavior of the MEP near the nanoparticle surface. Moreover, the height
of the first peaks in the counter-ion and co-ion density profiles
are of the same order but the former are located closer to the nanoparticle
surface dominating the contributions when integrated along the radial
separation distances.

A different scenario is predicted by eqn (\ref{eq:charging_mechanisim})
at higher pH levels (e.g. 6 to 8) where a large number of active functional
groups are deprotonated and very few are protonated, thereby inducing
large negative charge density on the nanoparticle surface. This behavior
generates strong nanoparticle-liquid electrostatic interactions and,
consequently, electrostatic energy contributions to the ionic PMF
comparable in magnitude with the corresponding ionic entropy energy.
In this case, unlike the results presented for low pH levels, the
ionic driving force near the nanoparticle surface strongly depends
on the ion species. For counter-ions specifically, we find that the
electrostatic potential energy competes with the ionic entropy (particle
crowding) energy by contributing both with positive values of the
same order to the corresponding ionic PMF. On the other hand, our
results for co-ions show that the electrostatic energy (negative values)
is partially compensated by the ionic entropy contribution (positive
values). This competition and balance behavior are depicted in Fig.
\ref{fig:ion-contributions-5A-1} where we represent the ionic entropy
contribution by dashed blue and cyan lines, the electrostatic potential
energy contribution by dashed light green and dark green lines, the
ion-ion correlation contribution by dashed red and magenta lines,
and the ionic potential of mean force by dashed black (counter-ionic)
and gray (co-ionic) lines. As a results of this interplay, the structural
properties of spherical EDLs in alkaline electrolyte solutions present
a significant increase in the accumulation of counter-ions ($Na^{+}$)
and depletion of co-ions ($Cl^{-}$) near the surface of the nanoparticle
with respect to those in the bulk phase. Our results also reveal a
more pronounced layering formation in counter-ions than co-ions. For
instance, Figs. \ref{fig:pH-effects-on-profiles-5A-1} (a)-(b) shows
5 times and 13 times higher accumulation of $Na^{+}$ in the first
shell than in the bulk solution at pH level 6 and 8, respectively.
Whereas our results for $Cl^{-}$ show 3.75 times and 1.5 times higher
accumulation at pH level 6 and 8, respectively. In addition, our results
reveal higher negative values and longer ranged asymptotic decay of
the MEP at high pH levels (see dashed red and black lines in Fig.
\ref{fig:MEP-SPM-1} (a)).

Certainly, the short-ranged nanoparticle-liquid entropic interactions
are weakened at intermediate separation distances, e.g between \textasciitilde{}
12Å and 18Å from the nanoparticle surface. Consequently, the MEP dominates
the asymptotic behavior of the ionic PMF (see Fig. \ref{fig:ion-contributions-5A-1})
as well as the pH effects on the structural and electrostatic properties
of electrolyte solutions. Accordingly, an increase of pH levels induces
large values of the PMF and generates slower asymptotic decay (see
Fig. \ref{fig:MEP-SPM-1} (a)). This in turn generates a casi-monotonic
decay behavior in the ion density profiles (see Fig. \ref{eq:ionprofilefinal}
(a)-(b)). As expected, all nanoparticle-liquid interactions and pH
effects on the electrolyte solution properties at longer separation
distances vanish recovering the well-known electrolyte bulk properties.
Overall, the EDL properties remain unchanged at pH level 10 and higher
(results not presented in this article) because the nanoparticle reaches
the charge saturation limit\citep{key-9,key-29}. 

\begin{figure}
\begin{centering}
\includegraphics[scale=0.4]{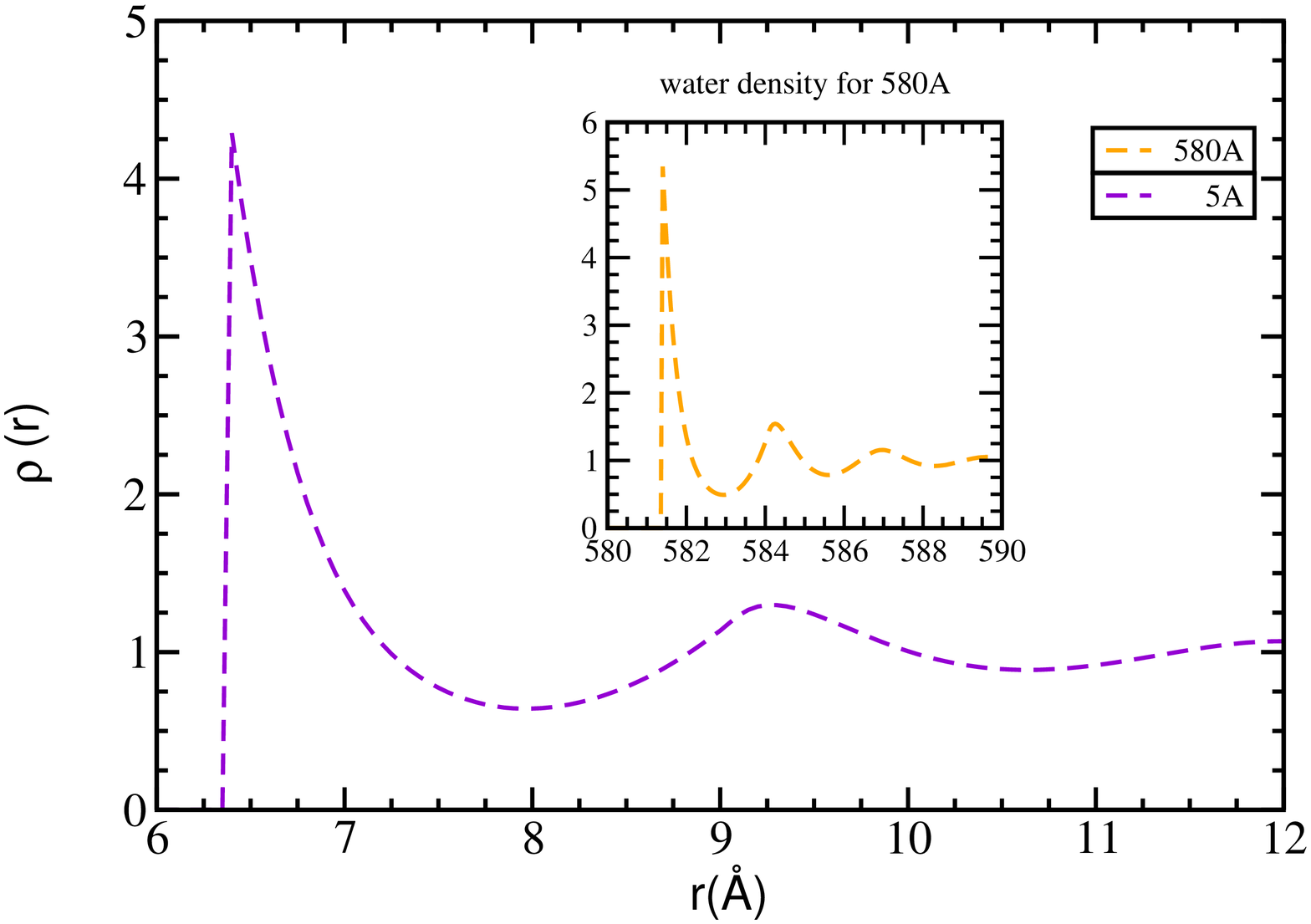}
\par\end{centering}

\caption{\label{fig:water-profile}Water density profiles per bulk concentration
predicted by CSDFT for 0.8$M$ $NaCl$ salt concentration and pH 4.
Purple and orange dashed lines correspond to 5Å ( and 580Å nanoparticle
sizes, respectively.}
\end{figure}

In our second analysis we repeat the previous calculations for a larger
nanoparticle size (= 580Å) in order to provide insight into the nanoparticle
size effects on the structural EDL properties of silica oxide nanoparticles.
Our results in Fig. \ref{fig:Hydrogen-profiles-SPM-1} (a) and (b)
show that the increase of the nanoparticle surface induces larger
number of deprotonated functional groups and, consequently, an increase
of the total nanoparticle charge. As a result, the corresponding values
of the MEP are magnified while keeping the trends predicted for the
smaller nanoparticle size (see Fig. \ref{fig:MEP-SPM-1} (a)-(b)).
Additionally, the increase of nanoparticle size causes higher entropic
interaction between the nanoparticle and the surrounding medium which
in turn increases the particle crowding effects on the ionic density
profiles (e.g. taller peaks) near the nanoparticle surface. As a result,
we find a more pronounced layering formation for the larger nanoparticle
size. For instance, we find around 25 \% more water molecules (see
Fig. \ref{fig:water-profile}), 122 \% more $Na^{+}$ ions and 33
\% less $Cl^{-}$ ions (see Fig. \ref{fig:pH-effects-on-profiles-5A-1}
(c)-(d)) accumulated in the first shell for 580Å than for 5Å nanoparticle
size at pH level 8. Overall, our results on the ion density profiles
and MEP show that the trends obtained on the EDL properties for a
5Å nanoparticle are magnified when the particle size is increased.


\subsection{PB - CSDFT comparison}

\begin{figure}
\subfloat[]{\includegraphics[scale=0.33]{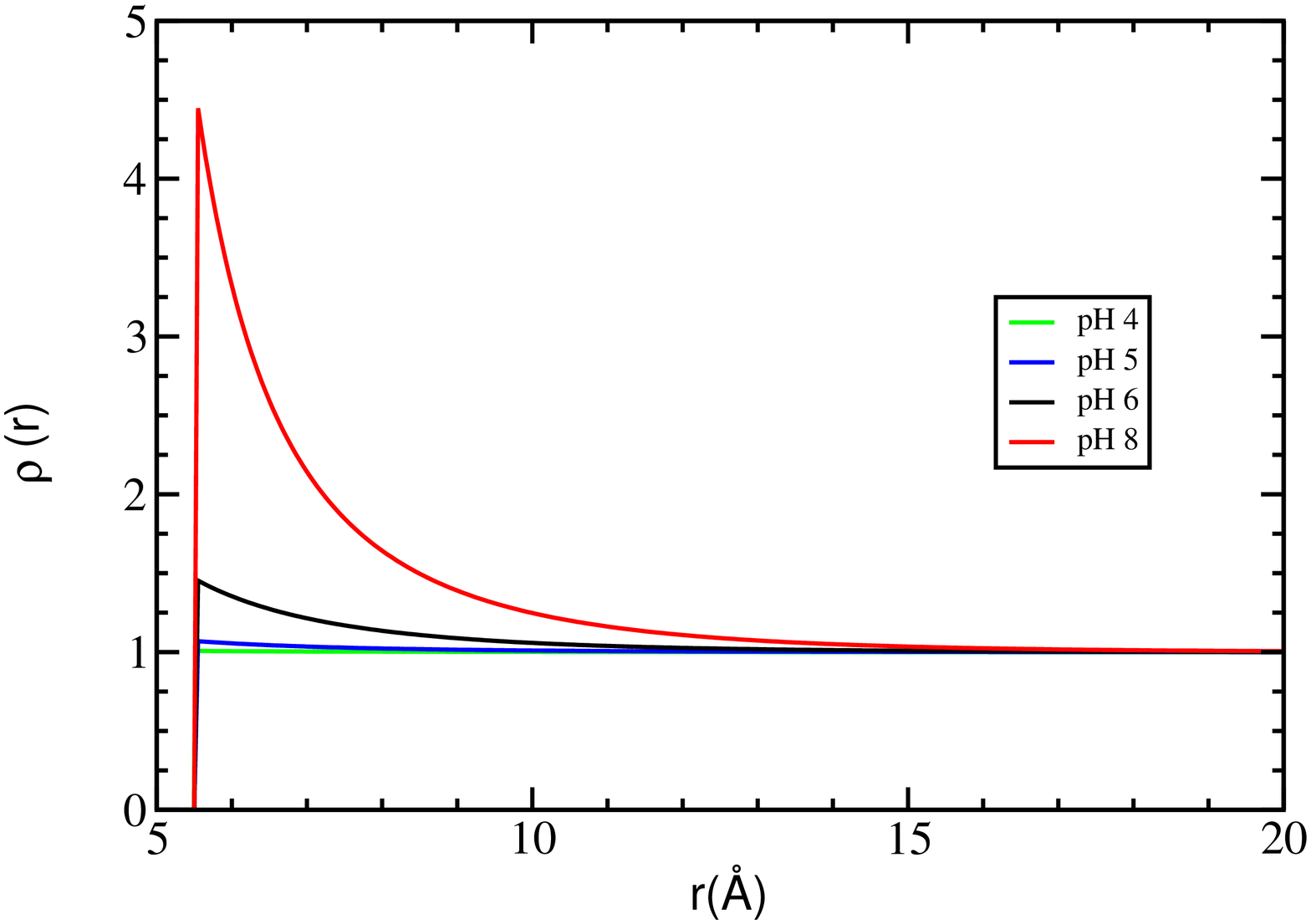}}\subfloat[]{\includegraphics[scale=0.33]{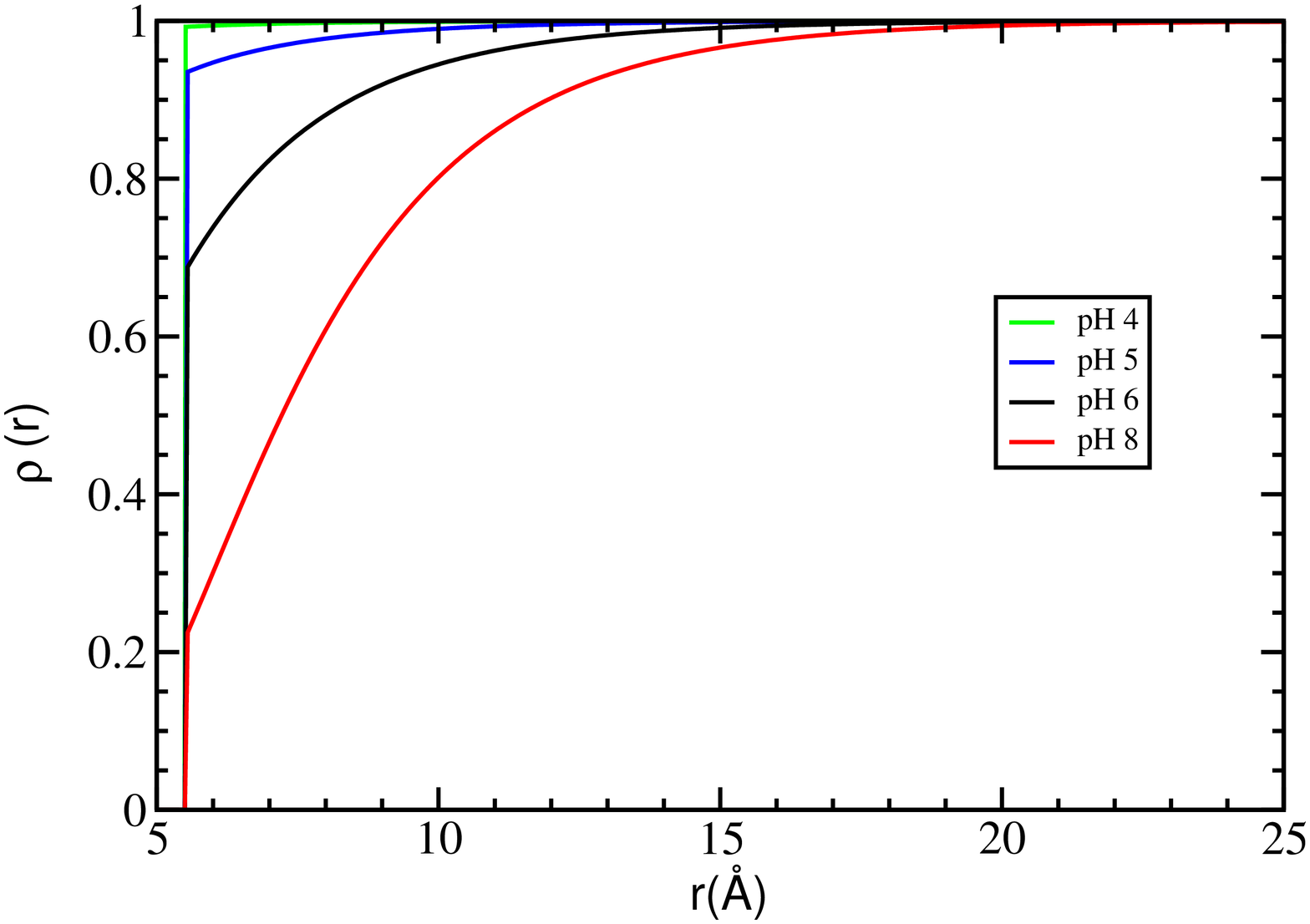}}

\subfloat[]{\includegraphics[scale=0.33]{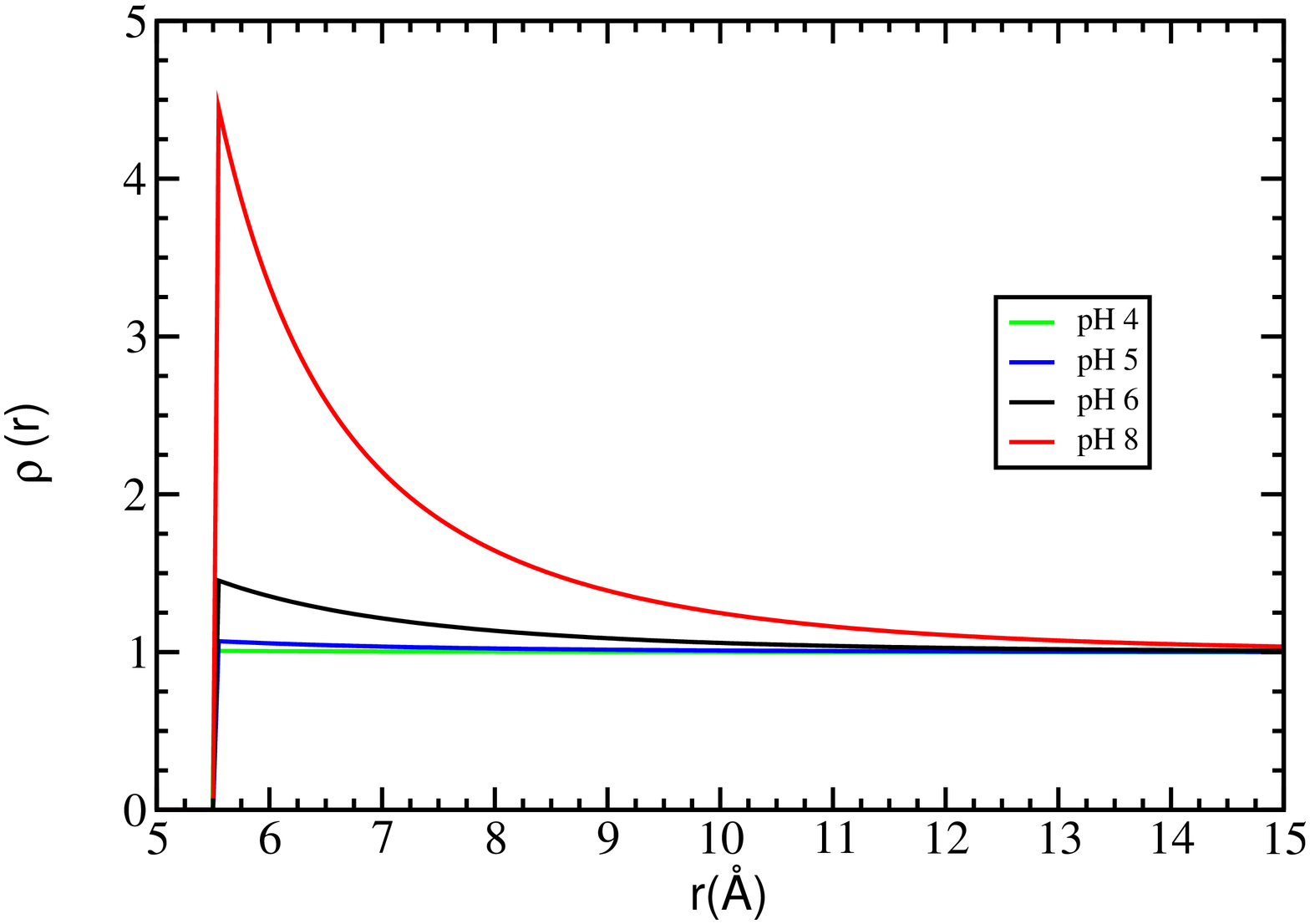}}\subfloat[]{\includegraphics[scale=0.33]{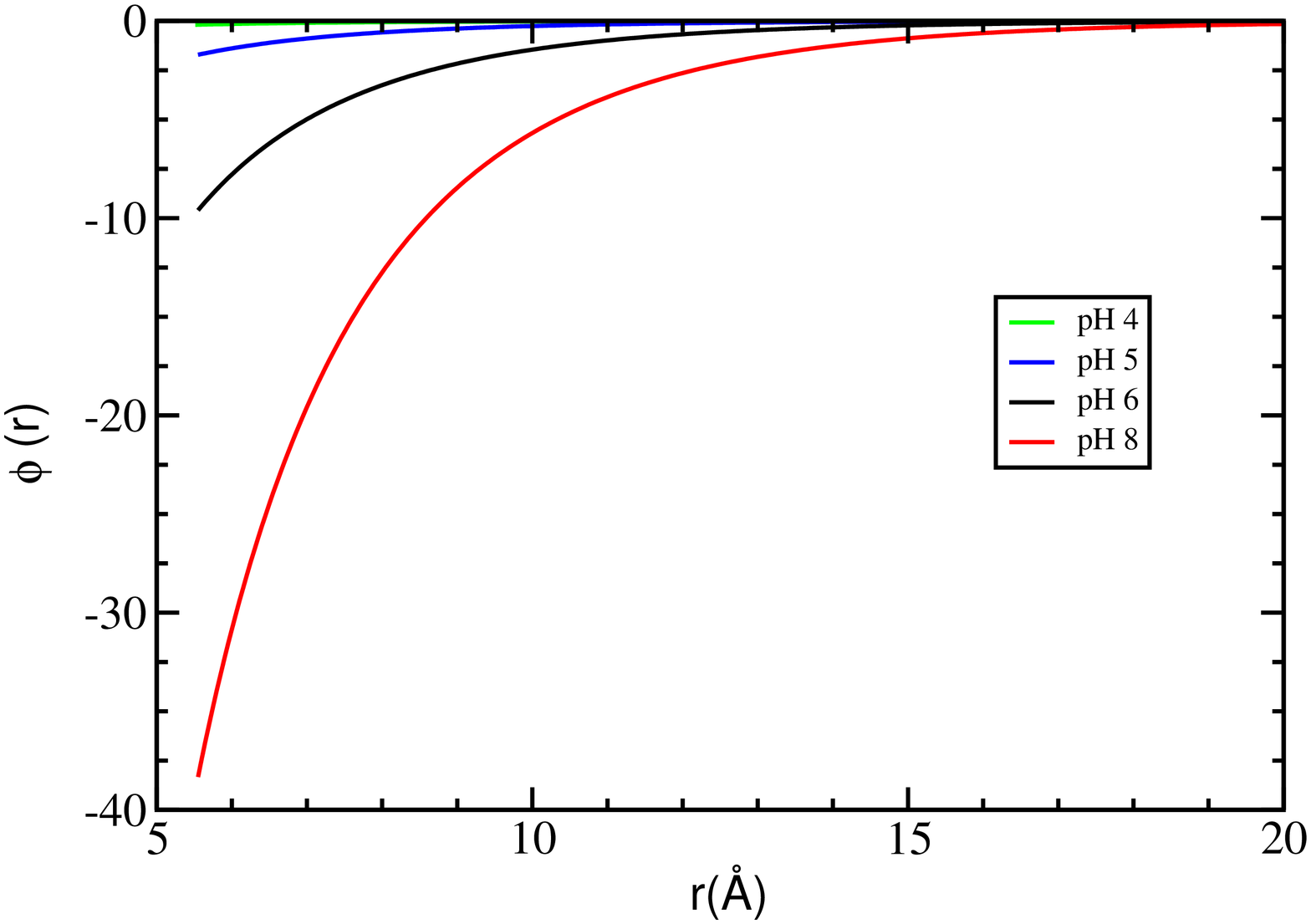}}\caption{\label{fig:PB-predictions-1}IPB results on 0.8$M$ $NaCl$ salt concentration
and 5Å nanoparticle size. Plots a, b and c correspond to $Na^{+}$,
$Cl^{-}$, and $H^{+}$ density profiles normalized by bulk concentration,
respectively. Plot d corresponds to the mean electrostatic potential
(MEP) per unit of thermal energy $KT$ Green, blue, black and red
solid lines represent the density profiles and MEP for pHs 4, 5, 6,
and 8, respectively. }
\end{figure}

In our final analysis we calculate PB predictions on ion density profiles
and MEP for 5Å nanoparticle size with the same electrolyte conditions
analyzed in the previous section. When comparing PB (solid lines in
Fig. \ref{fig:PB-predictions-1}) with CSDFT results (dashed lines
in Fig. \ref{fig:pH-effects-on-profiles-5A-1}-\ref{fig:MEP-SPM-1}),
we observed consistency in trends for the MEP and ionic density profiles
where long range contributions dominate, with significant deviations
near the nanoparticle surface, in which the short range effects of
the entropy and ion-ion correlation provide unneglectable corrections
to the PB predictions. These deviations are found to be more pronounced
at low pH levels where the balance between the number of deprotonated
and protonated functional groups on the nanoparticle surface significantly
attenuate the solute-liquid electrostatic interaction. Among the differences
between both approaches, it is worth mentioning that the PB approach
is not able to capture layering formation, charge inversion or co-ion
accumulation near the nanoparticle surface for any of the pH levels
and nanoparticle sizes considered in this work. Clearly, the source
of the differences is manifold. Firstly, the expression for the surface
charge layer position used in CSDFT $s\equiv R+N_{A}\lambda_{B}^{3}\sum_{i}z_{i}^{2}[\rho_{i}^{0}]d_{i}/(2m)$
generalizes the definition of continuum models ($s=R$) by accounting
for the ion diameters and bulk densities. As a result, it predicts
SCD layering formation closer to the nanoparticle surface at low ionic
densities for all ion sizes. On the other hand, it predicts longer
deviation from the nanoparticle surface at 0.8M electrolyte concentration.
This shift in the position of the SCD affects the evaluation of the
ZP ($\zeta\equiv\psi(r,\{[\rho_{j}]\})|_{r=s}$) generating lower
CSDFT values compared to those predicted by PB {[}{]}. Secondly, the
ZP predicted by CSDFT also accounts for the ion-ion correlations,
size asymmetry and particle crowding effects (see eqn. (\ref{fig:PB-CSDFT})),
which are omitted in continuum models. Additionally, these two factors
have a non-linear high impact effect on the titration and nanoparticle
SCD properties (see eqn (\ref{eq:complexation})). Another source
of difference corresponds to the expression used by CSDFT to estimate
the ionic PMF (see eqn (\ref{eq:ionprofilefinal})). Indeed, by considering
not only electrostatic but also entropy and ion-ion correlation interactions,
and in particular, by accounting for explicit water molecules at experimental
size and concentration, CSDFT has been capable of capturing ionic
layering formation and charge inversion, among other important features
of colloidal systems. On the contrary, these corrections to continuum
models are attenuated at lower electrolyte concentrations, obtaining
PB and CSDFT predictions in much better agreement\citep{key-1,key-9,key-29}.

\section{Conclusion}

In this article, we use classical solvation density functional theory
(CSDFT) and a surface complexation model to investigate the effects
of pH and nanoparticle size on the structural and electrostatic properties
of an electrolyte solution surrounding a spherical silica oxide nanoparticle.
This approach has shown to be an efficient and accurate computational
tool in accounting for mean electrostatic potential (MEP), ion-ion
correlation and ionic entropy (particle crowding) effects on surface
titration. Additionally, the formulation has been able to capture
the balance and competition between electrostatic and entropic contributions
to the total ionic potential of the mean force (PMF). This feature
has been particularly useful for identifying dominant interactions
governing nanoparticle electrical double layer properties under a
variety of pH levels and nanoparticle sizes. Overall, we find a high
impact of the pH level on the structural and electrostatic properties
of spherical electrical double layers in 5Å and 580Å nanoparticle
sizes immersed in 0.8$M$ monovalent electrolyte solution. At low
pH levels, our results show small number of charged functional groups
on the nanoparticle surface, and weak electrostatic interactions between
the nanoparticle and surrounding liquid. Therefore, the resulting
driving force governing the structural and electrostatic properties
of the (EDL) in acid electrolyte solutions is found to come from the
ionic entropy energy. The ion density profiles are characterized by
an increase in accumulation of co-ions ($Cl^{-}$) and depletion of
counterions ($Na^{+}$) near the surface of the nanoparticle, where
asymmetry size effects are responsible for more co-ions inhabiting
the first shell than counter-ions. Furthermore, our results on MEP
reveal charge inversions and a non-trivial, non-monotonic short-range
behavior. We find a different scenario at high pH levels where many
functional groups are deprotonated on the nanoparticle surface. In
this case, the ionic driving force near the nanoparticle surface is
found to depend heavily on the ion species. For counter-ions, the
electrostatic potential energy competes with the entropy, both providing
with positive values of the same order to the ionic PMF. However,
the electrostatic energy for co-ions is negative while entropy contributions
remain positive, displaying a balancing relationship between them.
As a result of this interplay, a reduced PMF governs the co-ion density
distribution. We find a significant increase in the accumulation of
counterions ($Na^{+}$) and depletion of co-ions ($Cl^{-}$) near
the surface of the nanoparticle in alkaline electrolyte solutions.
When comparing Poisson-Boltzmann (PB) with CSDFT results, we observed
consistency in trends for the MEP and ionic density profiles where
long range contributions dominate, with significant deviations near
the nanoparticle surface, in which the short range effects of the
hard sphere and ion-ion correlation provide unneglectable corrections
to the PB predictions. These deviations are found to be more pronounced
at low pH levels where the poor charging mechanism on the nanoparticle
surface significantly attenuate the solute-liquid electrostatic interaction.
Finally, our results on nanoparticle size effects show that increasing
the nanoparticle surface and keeping fixed the other parameters of
the model generates an increase of the nanoparticle charge and entropy
interaction with the electrolyte solution. Consequently, the trends
obtained on the EDL properties for a 5Å nanoparticle are magnified
when the particle size is increased. Further research is currently
in progress to study pH effects on cylindrical EDL properties. Future
work involves the extension of the polar solvation classical density
function theory, recently introduced by Drs. Varsasky and Marucho
for planar geometries, to study polarization and pH effects on cylindrical
and spherical EDLs.

\section*{Acknowledgments}

This work was partially supported by NIH Grant 1SC2GM112578-01.

\end{document}